\documentclass[%
 aip,
 amsmath,amssymb,
 reprint,%
]{revtex4-1}

\usepackage{graphicx}
\usepackage{dcolumn}
\usepackage{bm}

\usepackage[utf8]{inputenc}
\usepackage[T1]{fontenc}
\usepackage{mathptmx}
\usepackage{comment}
\usepackage{lineno}
\usepackage{xcolor}
\usepackage{verbatim}
\usepackage{float} 
\begin{document}

\preprint{AIP/123-QED}

\title[]{Drop spreading dynamics with a liquid needle drop deposition technique}
\author{Abrar Ahmed}
\author{Anuvrat Mishra}
 %
\thanks{The first two authors have equally contributed to the manuscript}
\author{MD Farhad Ismail}
\author{Aleksey Baldygin}
 \affiliation{ 
\textit{interfacial} Science and Surface Engineering Lab (\textit{i}SSELab), \\ Department of Mechanical Engineering,\\ University of Alberta,{ Edmonton, Alberta T6G2G8, Canada.}
}%

\author{Thomas Willers}
\affiliation{ KR\"USS GmbH, Borsteler Chaussee 85, 22453 Hamburg, Germany}

\author{Prashant R. Waghmare}
\email{waghmare@ualberta.ca}
\affiliation{ 
\textit{interfacial} Science and Surface Engineering Lab (\textit{i}SSELab), \\ Department of Mechanical Engineering,\\ University of Alberta,{ Edmonton, Alberta T6G2G8, Canada.}
}%

\date{\today}

\begin{abstract}
This paper represents a theoretical and an experimental study of the spreading dynamics of a liquid droplet, generated by a needle free deposition system called the liquid needle droplet deposition technique. This technique utilizes a continuous liquid jet generated from a pressurized dosing system which generates a liquid drop on a substrate to be characterized by optical contact angle measurements. Although many studies have explored the theoretical modelling of the droplet spreading scenario, a theoretical model representing the spreading dynamics of a droplet, generated by the jet impact and continuous addition of liquid mass, is yet to be addressed. In this study, we developed a theoretical model based on the overall energy balance approach which enables us to study on the physics of variation of droplet spreading under surrounding medium of various viscosities. The numerical solution of the non-linear ordinary differential equation has provided us the opportunity to comment on the variation of droplet spreading, as a function of Weber number ($We$), Reynolds number ($Re$) and Bond number ($Bo$) ranging from  0.5-3, 75-150, and 0.001-0.3, respectively. We have also presented a liquid jet impact model in order to predict the initial droplet diameter as an initial condition for the proposed governing equation. The model has been verified further with the experimental measurements and reasonable agreement has been observed. Experimental observations and theoretical investigations also highlight the precision, repeatability and wide range of the applicability of liquid needle drop deposition technique.   
\end{abstract}

\maketitle

\section{\label{sec:level1}Introduction}
The wetting characteristics of a liquid droplet over a solid surface has been studied extensively in different manners: numerical approaches simulate the transient flow field using computational fluid dynamics codes \cite{yong2018sliding,pasandideh1996capillary}, while the theoretical approaches \cite{madejski1976solidification} assume a velocity profile in the droplet and seek an analytical solution. In an ambient fluid medium, when a drop of liquid rests in  thermodynamic equilibrium on a solid substrate, the angle measured at the three-phase contact line is called the equilibrium contact angle and widely termed as "Young's angle" \cite{snoeijer2008microscopic}. For obtaining the Young's angle, the drop deposition method is crucially important as identified by Shuttleworth in the middle of nineteenths century~\cite{shuttleworth1948spreading} and numerous researchers~\cite{waghmare2014needle,waghmare2013drop,kwon2011rapid,yildirim2005analysis,qian2009micron}. Traditionally, the drop deposition is attained by bringing a pendant drop in close proximity to the substrate and allowed it to detach from the needle.  This detachment of the drop from the needle depends on several factors, such as needle and surface energy of the characterizing substrate~\cite{waghmare2013drop}, needle surface roughness\cite{mchale2004topography}, and drop volume~\cite{yuan2013contact}, retraction speed~\cite{qian2009micron} of the needle as well as operator's skills.  These factors become ordaining for very low or high energy surfaces and every aspect affects the resultant equilibrium contact angle. The drop weight method \cite{yildirim2005analysis} is another technique used mainly for low surface energy solids.  In such cases,  the drop pinch-off generates a capillary wave that  generates  significantly high pressure difference across the drop-medium interface~\cite{kwon2011rapid,kolinski2014drops}. In some cases, this pressure restricts the drop to spread on high surface energy surfaces as shown by Mahadevan et al.\cite{kolinski2014drops} where the water drop rebound scenario was reported on the glass substrate. 

There is a surge in the development of surfaces with known surface energy, in particularly the very low or very high energy surfaces and accurate contact angle measurement is one of the most reliable as well as most commonly used  methods to characterize the accurate surface energy of these novel substrates. Owing to the limitations of classical contact angle measurement techniques posed in such cases, it is essential to develop a needle-free technique to overcome such shortcomings~\cite{taberner2012needle}. A couple of novel techniques are developed to circumvent these limitations  namely needle-free drop deposition technique \cite{waghmare2014needle,waghmare2013drop, waghmare2015needle} and liquid-needle technique  \cite{jin2016replacing}. The needle-free drop deposition technique involves the selection of a separate low energy interface for the success of the technique hence it has very limited application whereas the liquid-needle surged as a promising alternative as universal needle-free technique. In the liquid-needle technique, a continuous liquid jet of micron size is used to generate the drop of size three to four orders of the jet size. 

 A theoretical model based on energy balance for droplet spreading must include inertial, viscous, gravitational and capillary forces. Madejski \cite{madejski1976solidification} perhaps was the first to report an overall energy balance (OEB) approach where the momentum with energy balance is utilized to quantify the maximum spread of a drop upon impact on a cold surface. Gu and Li \cite{gu2000liquid} extended OEB approach to model spontaneous spreading with relatively lower impact speeds (impact speed $<= 2.5 m/s$). Erickson et al. \cite{erickson2001energy} successfully deployed the OEB approach to study hydrodynamically driven forced spreading where droplet was generated by quasi-static addition of mass through a punctured surface. However, the major drawbacks of punctured surface is that while creating a hole in the surface, the surface properties may be hampered which as a result may affect the spontaneous spreading of the droplet. In addition, Erickson et al.\citep{erickson2001energy} also did not analyse the effect of outer medium on their hydrodynamic model of droplet spreading and the effect of boundary movement work. In the present study the liquid needle droplet deposition has been utilised where the drop is generated on the substrate by a laminar jet emanating from a pressurized dosing system above the substrate. The present work follows the OEB approach to develop a model that predicts drop spreading until it attains the equilibrium.

The prime motive of the present study is to identify the role of various factors which can affect the dynamics of drop spreading and identify the limitations of the liquid needle technique. A theoretical model is developed and validated with experimental results for drop spreading where a continuous jet adds the mass to the drop via  liquid-fluid interface~\cite{jin2016replacing}. An droplet impact study is also performed to predict the maximum spreading of the droplet immediately after the jet impact  \citep{chen2017drop,ahmed2018maximum,ahmed2018effects,guo2016investigation}. Finally, a non-dimensional analysis is performed to obtain a  phase plot that can comment on applicability of the liquid needle technique for a given drop-medium combination.

\section{\label{sec:level1}Theoretical model}
\begin{figure*}
 \centering
\includegraphics[scale=0.5]{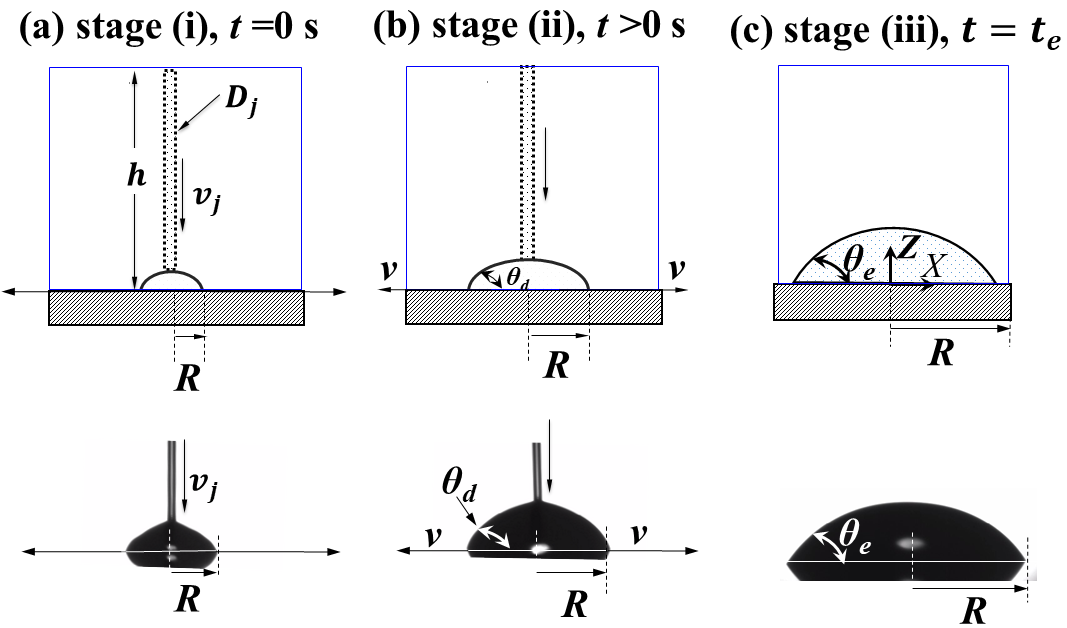}
\caption{ Different stages involved in liquid jet drop deposition technique. (a) Liquid jet with diameter $D_j$, density $\rho_d$ and viscosity $\mu_d$ impinging on a substrate from height $h$ at a velocity $V_j$ (stage (i), $t=0s$) in the presence of fluid medium with density $\rho_m$ and viscosity $\mu_m$, (b) drop growth with a radial velocity of $v$ (stage (ii), $t > 0s$) and dynamic advancing contact angle of $\theta_d$, (c) drop with equilibrium contact angle of $\theta_e$ after the ceasing of the jet (stage (iii), $t=t_e$). The top panel of the Figure represents the pictorial representation of each stage whereas the corresponding experimental observation is depicted in bottom panel.}
\label{Figure_1}
\end{figure*}

 The present work follows the OEB approach to develop a model that predicts drop spreading until it attains the equilibrium.  The diameter and momentum of the jet is carefully chosen to avoid any splashing when the liquid hits the solid substrate\cite{PatentPDS}. The jet on reaching the solid substrate is assumed to form a droplet with the shape of a spherical cap (Stage $- i$ ) and the jet continues to feed the liquid  (Stage $- ii$) until the desired volume is achieved (Stage $- iii$). Based on the dynamics occurring in the control volume this process can be trisected as depicted in Figure~\ref{Figure_1}. Considering the deformable drop, the energy imparted by the impinging jet is transformed into internal energy, surface energy, and gravitational energy, in addition to the resistance offered by the medium viscosity and viscous dissipation within the spreading droplet. The energy transferred from the impinging jet to the spreading drop results in Eq.~\ref{Eq1} as follows:
\begin{equation}
\dfrac{dE_{in}}{dt}= \dfrac{d}{dt}(E_{system}+E_s+E_g)+\dfrac{d}{dt}(W_{vd}+W_{mv}) 
\label{Eq1} 
\end{equation}
Here, $E_{in}$ is the incoming energy available in the liquid jet, $E_{system}$ is the internal energy in the spreading droplet, $E_s$ is the surface energy, $E_g$ is the gravitational potential energy, $W_{vd}$ is the work due to viscous dissipation and $W_{mv}$ is the work associated with medium viscosity. 

The energy available in the jet is the combination of enthalpy \cite{sonntag1998fundamentals} and jet kinetic energy. It has already been demonstrated with order of magnitude analysis that for isobaric and isothermal
conditions  the overall change in the enthalpy is negligible\cite{erickson2001energy}. The jet initial velocity $(v_{j0})$ reduces to $v_j$ due to the presence of the medium and can be quantified as,~\cite{soh2005entrainment} $v_j=v_{j0} \left[\dfrac{1}{1+\frac {\sqrt{C_D}}{{2}}}\right]$. Here, $C_D$ is the drag coefficient which is related with the jet frontal geometry.  Considering the infinitesimal change in mass with respect to time, the total rate of energy transfer is: 
\begin{equation}
\dfrac{dE_{in}}{dt}=\left[\dfrac{v_j^{2}}{2}\right]\dfrac{dm}{dt}
\label{Eq2}
\end{equation}

The internal kinetic energy induced due to the impingement of the jet on the liquid-medium interface can be ignored since the drop surface area is remarkably greater than the liquid jet cross sectional area \cite{erickson2001energy}. The  initial drop diameter ($D_0$) is large enough in comparison to the jet diameter ($D_j$) to ignore drop internal kinetic energy. The total surface energy of the system can be defined considering the surface energies of the three interfacial phases, $i.e.$, liquid-solid (drop-substrate), liquid-fluid (drop-medium) and solid-fluid (substrate-medium) which suggests $E_s= \sigma_{ds} A_{ds} - \sigma_{sm} A_{sm} +\sigma_{dm} A_{dm}$, where, $\sigma$ and $A $ represent the surface energy and area for respective interfaces and subscripts $d$, $s$ and $m$ denotes the drop, solid and the medium, respectively. With a spherical drop shape assumption the rate of change of surface energy is
\begin{equation}
\dfrac{dE_s}{dt}=2 \pi R \sigma_{dm}[2h(\theta_d)-\cos\theta_e] \dfrac{dR}{dt}
\label{Eq3}
\end{equation}
where, $h(\theta_d)= \dfrac{1-\cos\theta_d}{\sin^2 \theta_d}$
and $\theta_d$ and  $\theta_e$ are the advancing (dynamic) and equilibrium contact angle, respectively, as depicted in Figure~\ref{Figure_1}.  For an infinitesimal increase of mass, $\Delta m $, during the continuous growth of the drop, if the center of gravity alters by $\Delta z$, the rate of change in potential energy can be expressed as $\dfrac{dE_g}{dt}=g \left [m\dfrac{dz}{dt}+z\dfrac{dm}{dt} \right ] $ \cite{erickson2001energy} where the total mass of the system can be obtained as $m=m_0+\int\limits_a^b \dfrac{dm}{dt} dt$. 

The resulting rate of change in the gravitational potential energy can be expressed as:
\begin{equation}
\dfrac{dE_g}{dt}= g \left [m \left (\dfrac{3f(\theta_d)}{4} \dfrac{dR}{dt} \right)+ \dfrac{R}{4} f(\theta_d)\dfrac{dm}{dt} \right ]
\label{Eq4}
\end{equation}
where, 
$f(\theta_d)=\dfrac{2-\sin^2 \theta_d + 2\cos \theta_d}{(2+\cos\theta_d)\sin\theta_d}$.

Due to the continuous growth of the drop volume, the surrounding medium gets displaced that requires additional work which is considered as a resistance due to the surrounding medium viscosity. The role of medium viscosity can be ignored in case of air  as presented in the numerous studies \cite{madejski1976solidification, gu1998model} but for the case of liquid needle,the drop deposition in the medium other than air is also studied in detail where the medium viscosity becomes paramount to consider. The shear stress at the drop boundary, $i.e.$, at the liquid-medium interface, can be defined as 
$\tau=2 \mu_m \left (\dfrac{\partial u}{\partial R} \right )_{r=R}$
  , where, $\mu_m$ is the medium viscosity and $u$ is the velocity by which the surrounding medium is being displaced. If we consider a lamina of fluid outside the drop at a distance $r$, where $r>R$, and implement the mass conservation method, we  obtain the velocity of the surrounding medium adjacent to the drop boundary. The resultant rate of work due to medium viscosity can be  given as,
\begin{equation}
\dfrac{dw_{mv}}{dt}= \left [\dfrac{4 \mu_m}{R} \dfrac{dR}{dt}\dfrac{1}{\rho_m} \right ] \dfrac{dm}{dt}
\label{Eq5}
\end{equation}
where, $\rho_m$ is the surrounding medium density.

To describe the viscous dissipation during spontaneous sessile drop spreading, different models have been proposed, e.g., De Gennes \cite{brochard1992dynamics,de1985wetting} predicted the viscous dissipation work based on lubrication approximation whereas Chandra et al. \cite{chandra1991collision} suggested alternative approach  using a laminar boundary layer within the splat of the impacting drop.  For hydrodynamic drop spreading, we follow the De Gennes approach \cite{brochard1992dynamics,de1985wetting}  and observed a good agreement with the experimental observations in particularly  for the cases where hydrodynamic forces are remarkably larger than the molecular forces \cite{batchelor2000introduction}.
Based on the lubrication model \cite{brochard1992dynamics,de1985wetting}, the viscous force per unit length of the three-phase contact line can be expressed as, $ F_v= \dfrac{3 \mu_{d}}{\theta_d} ln \left (\varepsilon^{-1} \dfrac{dR}{dt} \right )$, where $\theta_d$ is the instantaneous dynamic contact angle, $\mu_{d}$ is the viscosity of the droplet and $\varepsilon$ is the ratio of the microscopic length $(L_\delta)$ to macroscopic cut-off length $(L)$. In general, $L_\delta$ may vary between 1$~\mu m$ to 5~$\mu m$ whereas $ L$ can be defined as the horizontal length scale $(R)$ of the drop \cite{erickson2001energy}. The viscous dissipation work of the circular three phase contact line is $2 \pi RF_v$.  Therefore, the viscous dissipation work per unit time over the three-phase contact line is~\cite{brochard1992dynamics,de1985wetting} 
\begin{equation}
\dfrac{dw_{vd}}{dt}=6 \pi \mu_d ln(\varepsilon^{-1})\dfrac{R}{\theta_D} \left (\dfrac{dR}{dt}\right )^2
\label{Eq6}
\end{equation}

As mentioned earlier, we also consider boundary layer approximation as an alternative approach to lubrication approximation, for predicting the viscous dissipation for droplet-substrate combination with a higher contact angle. For this case, we have adopted the boundary layer approximation model for a higher contact angle system, suggested by Chandra et al.~\citep{chandra1991collision}, which is also numerically verified by Guo et al.~\citep{guo2018spreading} for a wide range of viscosities. Based on the boundary layer approximation model the viscous dissipation work is approximated as, $w _{v}=\int_{a}^{b} \phi\Omega t_{c}$. \citep{chandra1991collision,jin2016replacing,ahmed2018maximum}, where $\phi$ is the viscous dissipation which can be approximated as,$ \phi=\mu_{d} \left(\dfrac{\partial v_i}{\partial x_k}+\dfrac{\partial v_k}{\partial x_i}\right)\dfrac{\partial v_i}{\partial x_k}=\mu_{d} v_{j}^2/\delta^2 $, where $\delta=\dfrac{2D_j}{\sqrt{Re}}$ is the characteristic length scale of the droplet, $\Omega=\pi R^2 \delta$ is the volume of the droplet and $t_c=h_j/v_j=k_{h_{j}} D_j/v_j$. Therefore, considering the boundary layer approximation, the work done due to viscous dissipation per unit time, over the three-phase contact line, can be expressed as,
\begin{equation}
\dfrac{dw_{vd}}{dt}= \mu_d v_j \pi k_{h_{j}}\sqrt{Re} R \dfrac{dR}{dt} 
\label{Eq7}
\end{equation}

While considering the lubrication approximation, combining Eqs.\ref{Eq1}-\ref{Eq6} the governing equation for dynamic spreading of the drop with liquid needle deposition can be written as the following, 
\begin{eqnarray}
6 \pi \mu_d ln(\varepsilon^{-1})\dfrac{R}{\theta_D}  (\dfrac{dR}{dt} )^2  +  [2 \pi R \sigma_{dm}(2h(\theta_d)-\cos\theta_e)   \nonumber\\
 +  (m_0+\dfrac{dm}{dt}  )g \dfrac{f(\theta_d)}{4} -\dfrac{4 \mu_m}{R \rho_m} \dfrac{dm}{dt}  ] \dfrac{dR}{dt} \nonumber\\ +
 \frac{dm}{dt}[\frac{gRf(\theta_d)}{4}-\frac{v_j^2}{2}] =0
\label{Eq8}
\end{eqnarray}

The non-dimensionalized equation can also be obtained as depicted in Eq.~\ref{Eq9} where the characteristic length and velocity are considered as the jet radius and velocity, respectively. 
\begin{eqnarray}
\dfrac{6 ln(\varepsilon^{-1})}{\theta_d} \dfrac{R^*}{Re} \left (\dfrac{dR^*}{dt^*} \right )^2 \nonumber\\ +  \bigg[\dfrac{4R^*}{We} (2h(\theta_d)-\cos\theta_e)+ \nonumber\\
\dfrac{f(\theta_d)G(\theta_d)}{24} (R_{0}^{*})^3\dfrac{Bo}{We} + \dfrac{k_{h_{j}} f(\theta_d)}{4} t^* \dfrac{Bo}{We} \nonumber\\
+ \dfrac{k_{\mu_{m}}}{R^*} \dfrac{8}{Re} \bigg] \dfrac{dR^*}{dt^*}+\frac{dm}{dt} \bigg[\frac{f(\theta_d)}{4}R^* \frac{Bo}{We} \bigg] =0 
\label{Eq9}
\end{eqnarray}

Here, $Re=\rho_d v_j D_j / \mu_d$  , $We= \rho_d v_{j}^2 D_j / \sigma_{dm}$ , and $Bo = \rho_d g D_{j}^2/\sigma_{dm} $ are Reynolds number, Weber number and Bond number, respectively; also, $k_{\mu_{m}}=\mu_m / \mu_d$, $R^*=\dfrac{R}{D_j/2}$, $R_{0}^{*}= \dfrac{R_0}{D_j/2}$, $t^* = \dfrac{t}{D_{j}/v_{j}}$ and
 $G(\theta_d)=\dfrac{2-3\cos(\theta_d)+\cos^3(\theta_d)}{\sin^3(\theta_d)}$.\\

Similarly, for boundary layer approximation, combining Eqs.\ref{Eq1}-\ref{Eq5} and \ref{Eq7} the governing equation for dynamic spreading of the drop with liquid needle deposition can be expressed as the following,
\begin{eqnarray}
\mu_d v_j \pi k_{h_{j}}\sqrt{Re} R \dfrac{dR}{dt}  +  [2 \pi R \sigma_{dm}(2h(\theta_d)-\cos\theta_e)   \nonumber\\
 +  (m_0+\dfrac{dm}{dt}  )g \dfrac{f(\theta_d)}{4} -\dfrac{4 \mu_m}{R \rho_m} \dfrac{dm}{dt}  ] \dfrac{dR}{dt} \nonumber\\ +
 [\frac{gRf(\theta_d)}{4}-\frac{v_j^2}{2}\frac{dm}{dt}] =0
\label{Eq10}
\end{eqnarray} 
The non-dimensional form of Eq.~\ref{Eq10} can also be written as,
\begin{eqnarray}
\dfrac{k_{h_{j}}}{4\sqrt{Re}} R^* \dfrac{dR^*}{dt^*} +  \bigg[\dfrac{4R^*}{We} (2h(\theta_d)-\cos\theta_e)+ \nonumber\\
\dfrac{f(\theta_d)G(\theta_d)}{24} (R_{0}^{*})^3\dfrac{Bo}{We} + \dfrac{k_{h_{j}} f(\theta_d)}{4} t^* \dfrac{Bo}{We} \nonumber\\
+ \dfrac{k_{\mu_{m}}}{R^*} \dfrac{8}{Re} \bigg] \dfrac{dR^*}{dt^*}+\frac{dm}{dt} \bigg[\frac{f(\theta_d)}{4}R^* \frac{Bo}{We} \bigg] =0 
\label{Eq11}
\end{eqnarray}

The hydrodynamic spreading of a drop can be obtained theoretically by solving any of the equations from Eqs.~\ref{Eq8} to ~\ref{Eq12} with appropriate initial conditions. One of the initial conditions is the drop diameter at the first instance which can be calculated with the knowledge of droplet impact analysis. In traditional drop impact analysis, the splat shape of the drop is considered as the maximum spread condition\cite{chandra1991collision}. 
However, in case of an impacting jet, we can assume that at the first instant, the splat-shape is obtained with a drop volume equivalent to initial jet volume immediately before impact. 

To determine the initial spreading diameter of the splat we can further employ the energy balance equation, $i.e.$ the energy available in the jet (kinetic energy) before deposition and the energy transferred (surface energy, viscous dissipation work and work done due to medium viscosity) to form the splat droplet shape (initial drop shape). We can define the surface energy considering the splat shape as, 
$\frac{\pi}{4} D_0^{2} \sigma_{dm} (1- \cos \theta_e)$.   Considering De Genne's approximation \cite{brochard1992dynamics,de1985wetting} the viscous dissipation work can be calculated as, $6 \pi \mu_d ln(\varepsilon^{-1})\dfrac{R}{\theta_d} \left (\dfrac{dR}{dt}\right )^2$. Here, we can approximate the change in radius and time as $dR\approx D_0/2$ and $dt\approx t_c$. 
The time required for the droplet to form splat shape $(t_c)$ can be readily used from the traditional droplet impact analysis \cite{pasandideh1996capillary}.  Finally, the work due to medium viscosity considering the elemental area of the splat can be approximated as $\dfrac{3\pi k_{\mu_{M}}}{8} \dfrac{k_{h_{j}} D_0^{3} We}{D_j Re}$.

Thus, the non-dimensional equation for the initial spreading ratio $(\xi=D_0/D_j )$ is: 
 \begin{equation}
\xi^3 \left [\frac{9 ln(\varepsilon^{-1})}{32 \theta_d} + \dfrac{3 k_{\mu_{M}} k_{h_{j}}}{8} \right] \frac{We}{Re} + \frac{\xi^2}{4} [1-\cos \theta_e] - \frac{k_{h_{j}}}{8}We - k_{h_{j}}=0
\label{Eq12}
 \end{equation}

Again, by replacing only the viscous dissipation work, approximated by lubrication model, with the boundary layer approximation, we can write the non-dimensional energy balance equation for the splat formation as,\begin{equation}
\dfrac{1}{4} \dfrac{k_{\mu_{M}}}{k_{h_{j}}} \dfrac{We}{Re} \xi^3+\left [\dfrac{We}{8\sqrt{Re}}+\dfrac{1}{4}(1-cos(\theta_e)) \right]\xi^2 -\dfrac{k_{h_{j}}}{8}We=0
\label{Eq13}
 \end{equation}
 This equation \ref{Eq13} can be used as the initial condition for spreading diameter for the governing equation approximated with the boundary layer.

 On the other hand, we can use the continuity equation between the cross sectional diameter of the jet and the wetted area of the initially formed droplet to calculate the initial velocity with respect to known jet velocity, which can be used as another input parameter for the initial condition.
 \\ 
 \begin{figure*}
 	\includegraphics[scale=0.65]{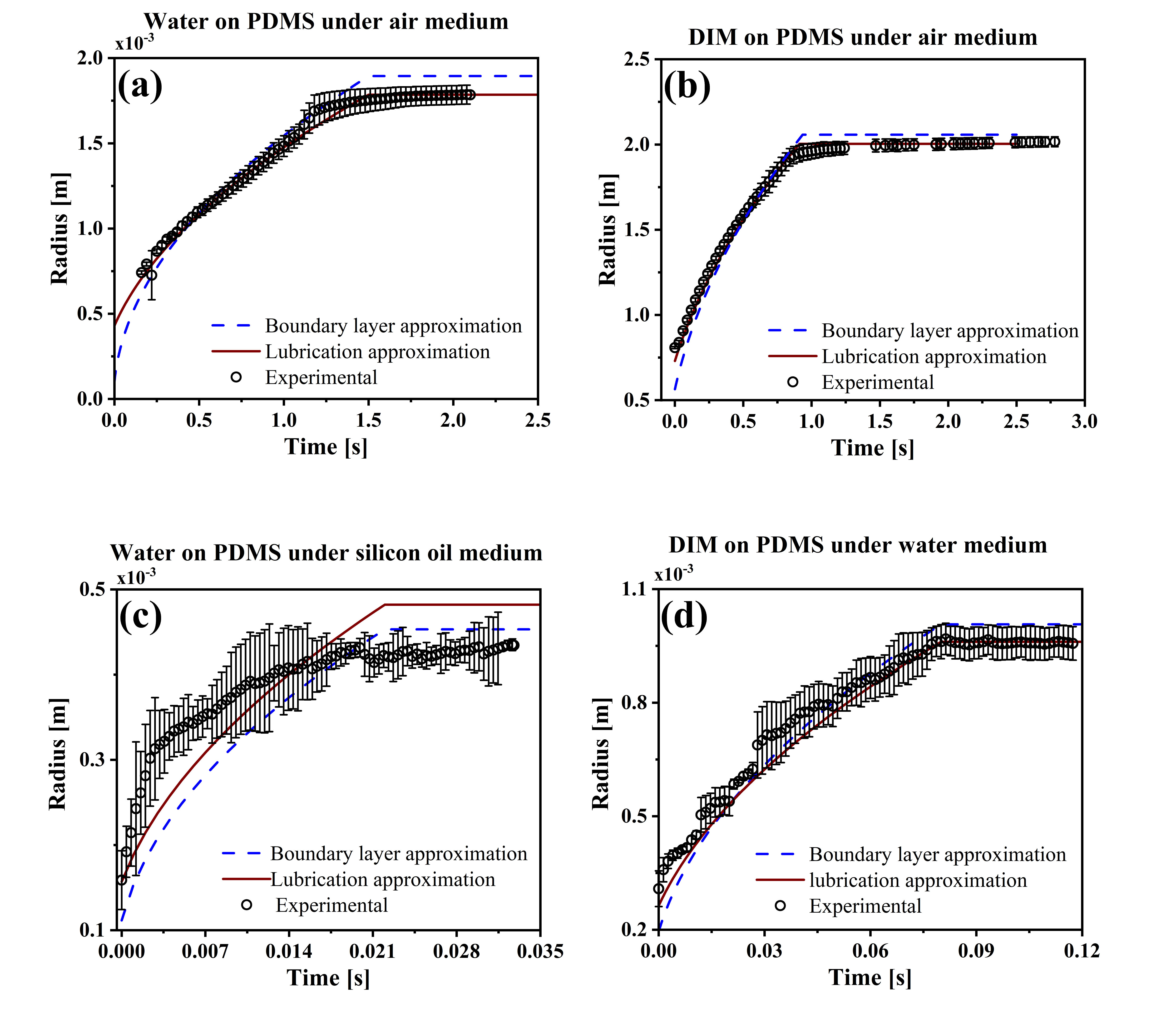}
 	\begin{center}
 	
 	\end{center}\centering
 	\caption{ Comparison of the measured drop growth (radius) with the theoretical analysis applying liquid needle drop deposition technique, (a) Water on PDMS under air medium, (b) DIM on PDMS under air medium, (c) Water on PDMS under Silicon oil medium, (d)DIM on PDMS under water medium }
 	\label{Fig_3}
 \end{figure*}
 
\begin{figure*}[t]
	\includegraphics[scale=0.55]{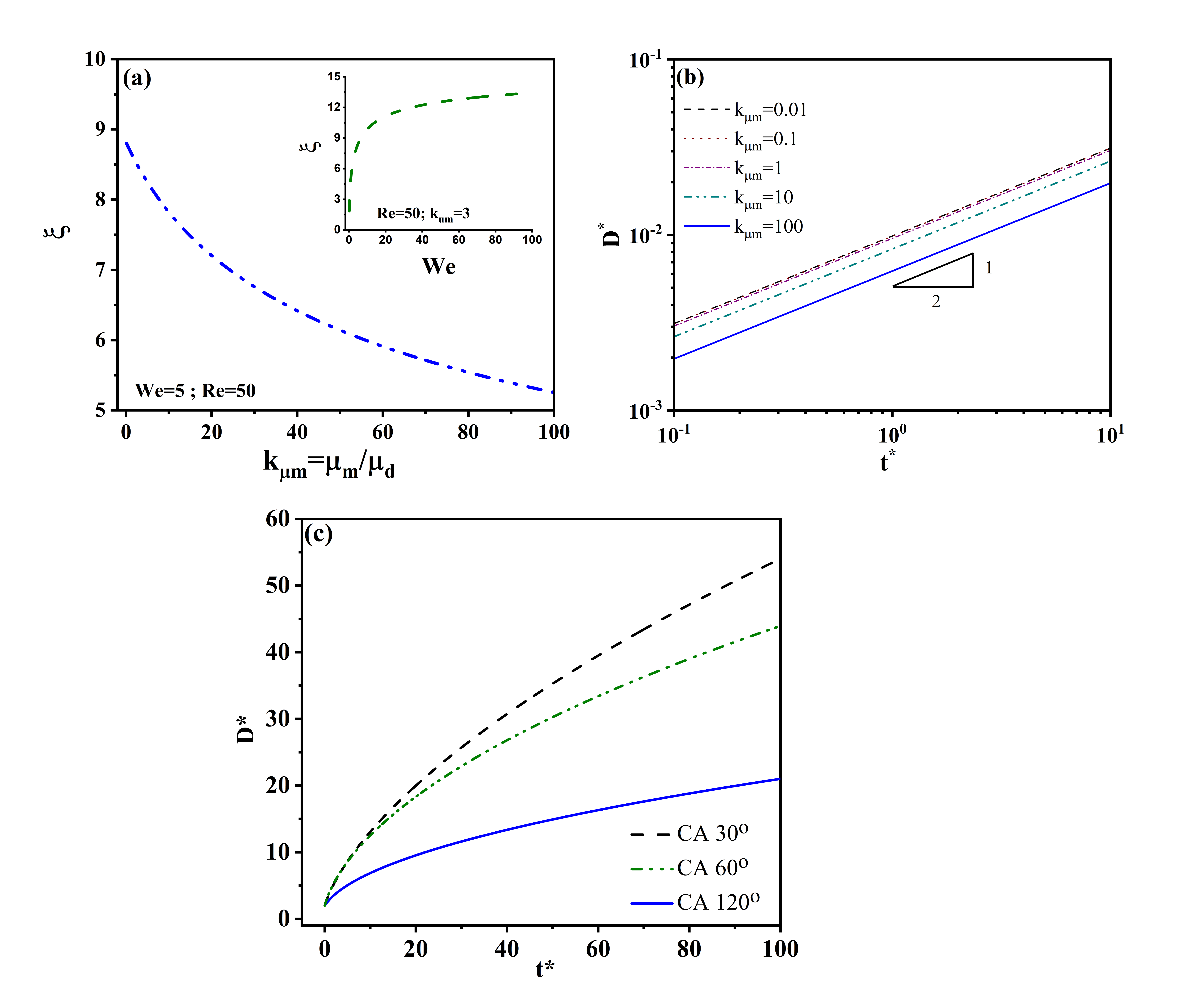}
	\centering
	\caption{Effect of viscosity ratio and contact angle on the spreading of the droplet deposited by a jet based deposition technique. (a) Variation of spreading ratio with respect to viscosity ratio for a fixed value of Weber and Reynolds number, whereas the inset Figure shows the relation between spreading ratio and the Weber number.(b)log-log plot for variation of nondimensional base diameter with respect to nondimensional time. (c) Transient variation of base diameter of a droplet for varying contact angle predicted by the model. }
	\label{Fig_5}
\end{figure*}

\section{\label{sec:level1}Results and Discussion}
The theoretical model is compared with experimental measurements performed in an air, water and silicon oil medium for different liquid jet and medium combinations. The experimental setup and methods used for this study is identical to the setup used previously by Jin et al. \cite{jin2016replacing}. The governing equation for the liquid needle, $i.e.$, Eq. \ref{Eq8}, has been numerically solved using the classical Runge-Kutta method (RK-4) \cite{harris1998handbook} for the drop base radius or diameter $(D)$. It is to be noted that while solving the equation, we assume that the dynamic contact angle $\theta_d$, equilibrium contact angle $(\theta_e)$ and the mass flux are known and the dynamic contact angle remains constant during  the entire drop deposition process. We have also varied the liquid-surrounding medium combinations having different viscosities and densities, e.g., water on polydimethylsiloxane (PDMS) in air medium, diidomethane (DIM) on PDMS in air medium, water on PDMS under silicon oil medium and DIM on PDMS under water medium which are shown in supplementary videos S1, S2, S3 and S4, respectively. For this particular study, the corresponding range of nondimensional parameters is considered as follows: $Re= 75-150$, $We =0.5-3$, $Bo= 0.001-0.3$, to avoid any unwarranted jet break-up\citep{yang2019manipulation}.

Figure~\ref{Fig_3} represents the comparison of experimental results with both models, lubrication and boundary layer approximation models, presented in this study. From Figure~\ref{Fig_3}, it can be seen that there is a marginal difference between these two models suggesting that each model has its own advantages and disadvantages. However, in most cases studied here, the lubrication model excels in predicting the droplet spreading except silicon oil medium. Boundary layer model can predict the water droplet spreading inside high viscous silicon oil medium better than the lubrication model. This result is consistent with our theoretical prediction, as inside the silicon oil medium, the water droplet contact angle is higher which is  $ \thicksim 148^{\circ}$. It is noteworthy to mention that while using the boundary layer model we have to use the appropriate value for viscous dissipation term which can be found in available literature \citep{mao1997spread,ahmed2018maximum}  , whereas in the lubrication model, the capillary cut-off length, $\epsilon$, is the only adjustable parameter.

Figure ~\ref{Fig_3} depicts the information of drop growth and spreading over time, where the solid and dotted line represents the theoretical model based on lubrication and boundary layer approximation, respectively whereas, symbols indicate the experimental outcome. From Figure ~\ref{Fig_3} it is evident that the theoretical model can successfully predict the transient variation of the spreading of the base radius except in one condition, which can be seen in Figure ~\ref{Fig_3}c. In case of Figure ~\ref{Fig_3}a, b, and d, the theoretical outcome is following the experimental observations. However, a relatively higher deviation, than the other medium, is observed in Figure ~\ref{Fig_3}c, representing the water droplet spreading under silicon oil medium. In the proposed model with both the approximations, the viscous dissipation inside the droplet is scrutinized, similarly, the viscous dissipation due to surrounding medium must be studied which is ignored in the proposed model. We suspect this might be the prime reason for the observed deviation for comparison in the case of liquid medium. Moreover, from Figure ~\ref{Fig_3}c we can see that experimentally the droplet stops spreading at $\thicksim$0.02s when the base radius is $\thicksim$0.35mm, whereas, the theory predicts the droplet of $\thicksim$0.45mm at the same stoppage. Therefore, it is evident that both models under predict additional energy dissipation due to surrounding medium and the considered boundary work needs to be supplemented by this additional medium dissipation for the closer prediction.  We also attribute this discrepancy to the loss of kinetic energy in liquid jet due to surrounding medium before impacting.\\

The role of the medium viscosity is further scrutinized by performing non-dimensional analysis which is presented in Figure ~\ref{Fig_5}.
Figure~\ref{Fig_5}(a) represents the variation of maximum spreading ratio, $\xi$ with respect to viscosity ratio, $k_{\mu_{m}}$ for a given $We$ and $Re$.  As suspected, the $\xi$ is decreasing gradually with respect to $k_{\mu_{m}}$, which supports the fact that surrounding medium suppress the spreading.  Increase in $k_{\mu_{m}}$ implies that the viscosity of surrounding medium is increased and thus corresponding energy loss to overcome the dissipation caused by the medium at the drop-medium as well as drop-substrate interfaces. Therefore, the $\xi$ decreases with the increasing of $k_{\mu_{m}}$ for a given $Re$ and $We$ scenario. The $ We$ considered in this study is a function of impact velocity which is the manifestation of jet velocity as well as the distance between the nozzle and the substrate. Higher the $We$, higher the impact velocity, which further assists droplet to spread. This phenomenon resembles with the inset of Figure ~\ref{Fig_5}(a), with enhancement in the input kinetic energy of the jet, the maximum initial spread increases accordingly. Surprisingly beyond certain $We$, for a given operating parameters $\xi$ remains almost the same and this motivated us to study the $Re$ variations as opposed to $We$. The plateau can be witnessed in the both the cases which resembles the same observation as presented by Clanet et al.\citep{clanet2004maximal}. It suggests that the $\xi$ is going through a transition from capillary regime to the viscous regime while varying with $We$. \\

In order to further analyse, the effect of viscosity ratio on the overall spreading of the droplet we have presented transient variation of total droplet spreading with respect to different viscosity ratio in Figure ~\ref{Fig_5}b. From Figure ~\ref{Fig_5}b, it is evident that for a fixed $We$ and $Re$, the spreading diameter of the droplet maintains a power law relationship of power index, $1/2$, with respect to time. However, it is also evident from the Figure ~\ref{Fig_5}b that viscosity ratio is not effecting the transient variation, rather it just shifts the magnitude of the droplet diameter. All the curves are collapsed on each other until the $k_{\mu_{m}}=1$, however change in droplet diameter is observed  when $k_{\mu_{m}}>1$. This implies as long as medium viscosity is lower than droplet viscosity, no significant change in droplet diameter can be observed for any fixed We and Re.\\

Figure ~\ref{Fig_5} (c) illustrates that drops with lower equilibrium contact angle, $i.e$ on surfaces with higher wettability, spreads faster compared to drops on substrates with higher contact angle, $i.e$ lower surface wettability. It is noteworthy to mention that the role of stick-slip motion at the contact line, which is always present in the case of air medium scenario, cannot be circumvented. Fluctuation in the instantaneous contact angle are caused due to such stick-slip motion and other numerous phenomena at contact line and  the proposed model requires further modifications to account these effects. Therefore, theoretical results are not compared to the experimental results.\\

\begin{figure}
	\centering
	\includegraphics[height=6.5 cm]{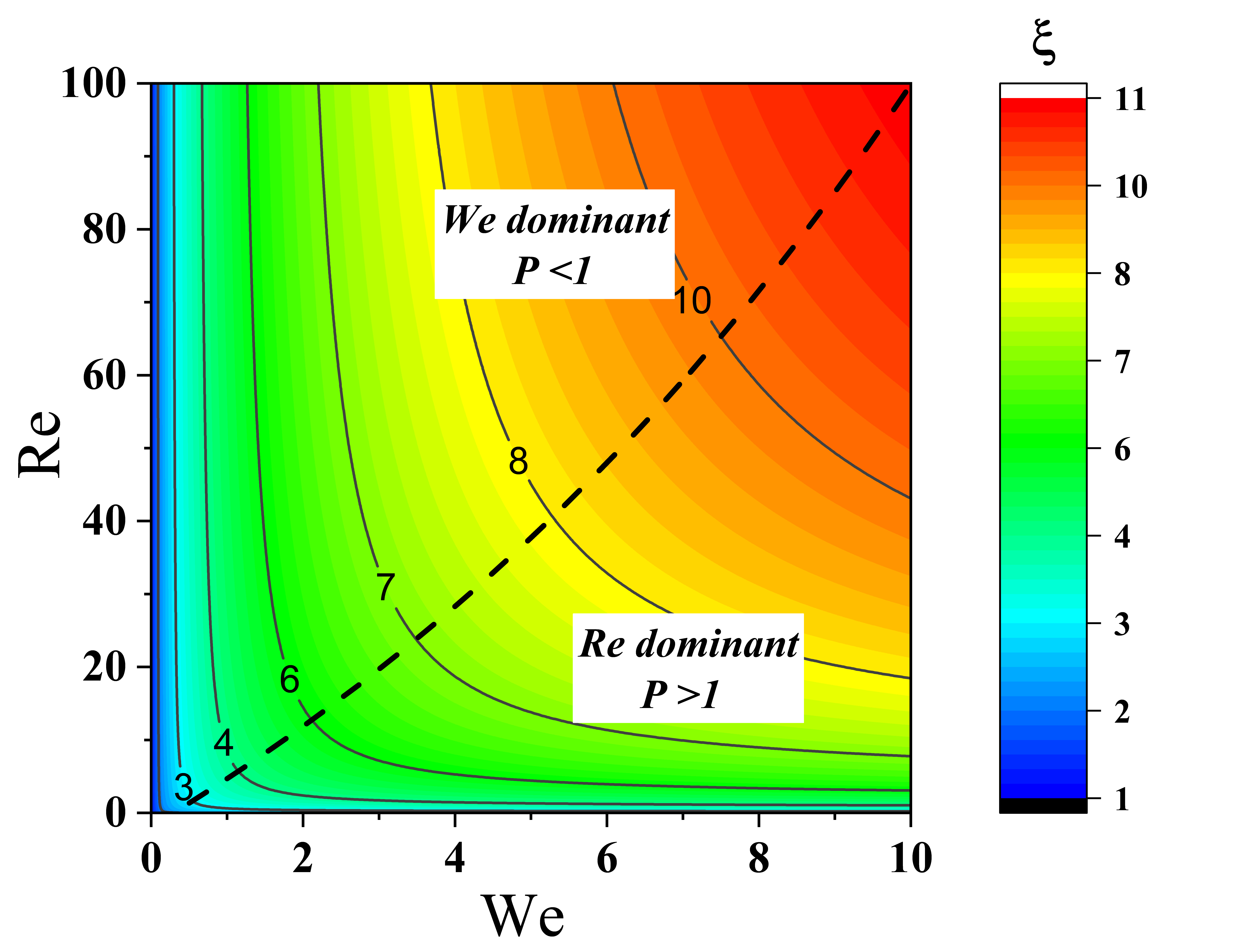}
	\caption{Phase plot showing the variation of maximum spreading ratio, $\xi$ with Reynolds and Weber number. The dotted curve divided the curve into two region: $ Re$ dominant and $We$ dominant region, whereas the solid lines represent the constant spreading ratio. The colour bar represents different values of spreading ratio, $\xi$. }
	\label{Fig_6}
\end{figure}

A parametric study was conducted to determine the limitations of the hydrodynamic model of droplet spreading process. From this study, a phase plot was produced as shown in Figure.~\ref{Fig_6}. This contour plot shows the variation of  $\xi$ (the initial maximum drop spreading in comparison with the jet diameter) with respect to various Reynolds and Weber numbers. Here the dotted curved line depicts the approximate inflection points for a constant $\xi$ curve. It denotes the boundary between the $Re$ and $We$ dominant flow regimes. \\ 

From Figure ~\ref{Fig_6}, it is evident that the maximum spreading diameter of an impacting jet is proportional to both Re and We. However, with respect to the sensitivity analysis, the Figure ~\ref{Fig_6} can be divided into two regimes – Re dominant and We dominant. Interestingly, for the Re dominant regime the spreading ratio is sensitive towards the change in Re, although the magnitude of Re is lower here compared to the corresponding We. The similar observations can be witnessed in the case of We dominant regime. For low viscous droplet the maximum spreading, $D_{max}$ can be scaled as\citep{clanet2004maximal} $D_{max} \sim D_0 We^{1/4}$ , whereas, for high viscous droplet the spreading is limited by the viscosity of the droplet, which yields another scaling law\citep{clanet2004maximal}, $D_{max} \sim D_0 Re^{1/5}$. Therefore, in order to explain the spreading scenario for a wide range of viscosity , Clanet et al.\citep{clanet2004maximal} defined a new nondimensional parameter as impact number, $P=We/Re^{4/5}$.  According to the study performed by Clanet et al. \citep{clanet2004maximal}, there is a transition of spreading ratio with respect to the impact number, $ P$. When, $ P<1$, the spreading is dominated by the capillarity regardless of the viscosity of the droplet, which yields a sharp increase in the maximum spreading ratio. However, when $P>1$, we can see that spreading ratio is independent of impact number, $P$ which implies the maximum spreading diameter of the impacting jet is significantly limited by the viscosity of the droplet. The transition from capillary to viscous regime takes place when the impact number, $P=1$. According to the definition of impact number, P, it can also be expressed as, $P=We/Re^{4/5}= \rho^{1/5} D_0^{1/5} v^{6/5} \mu^{4/5} \sigma^{-1}$.  From the definition of impact number it is to be noted that capillary regime, $P<1$, can be observed at small velocities, small viscosities and large surface tension, otherwise viscous regime will dominate the maximum spreading diameter. 

\section{\label{sec:level1}Conclusion}
This study presents the theoretical and experimental analysis of droplet spreading on a solid substrate for drops being deposited with the liquid needle dosing technique. In the liquid needle dosing system the droplet is formed by the continuous addition of mass through a liquid jet. The theoretical model for forced droplet spreading is developed based on an overall energy balance (OEB) equation. Further, the theoretical model is validated with the experimental results which show that the OEB model can successfully predict the transient droplet growth. The model also accounts for the surrounding medium with different viscosities and densities as according to both the theoretical and experimental analysis we can observe the spreading rate is hindered by the additional resistance of the surrounding medium viscosity. This study illustrates why the liquid needle dosing technique, adjusted with the appropriate parameters, can be the most careful drop deposition technique used for optical contact angle measurements. This is particularly interesting as it is also significantly faster as most other commonly used drop deposition techniques. This study provides next to the first experimental studies \cite{jin2016replacing}, a further more theoretical and experimental confidence level for the use of liquid needle drop deposition technique and also comments on the operating limitations of this technique. 

\section*{Supplementary Material}
The supplementary material provided as video files represents the droplet generation and spreading for different liquid-surrounding medium combinations having different viscosities and densities. Supplementary videos S1, S2, S3 and S4 represent water drop on PDMS in air medium, diidomethane (DIM) drop on PDMS in air medium, water drop on PDMS under silicon oil medium and DIM on PDMS under water medium , respectively.

\begin{acknowledgements}
The authors thank the Natural Sciences and Engineering Research Council (NSERC) for the financial support in the form of Grant No. RGPIN-2015-06542. As a part of the University of Alberta's Future Energy Systems research initiative, this research was made possible in part thanks to funding from the Canada First Research Excellence Fund.

\end{acknowledgements}

\section*{References }
\nocite{*}
\bibliography{aipsamp}

\end{document}